\documentclass[10pt]{iopart}

\usepackage{graphicx}  % needed for figures
\usepackage{dcolumn}   % needed for some tables
\usepackage{bm}        % for math
\usepackage{verbatim}   % for math
\usepackage{upgreek}   % for math
\usepackage{braket}
\usepackage{cite}
\usepackage{lineno}
\usepackage[load-configurations=abbreviations, redefine-symbols=true]{siunitx}
\usepackage{tikz}
\expandafter\let\csname equation*\endcsname\relax
\expandafter\let\csname endequation*\endcsname\relax
\usepackage{amsmath}
\usepackage{booktabs}
\usepackage{soul}
\usepackage{tablefootnote}
\usepackage[symbol]{footmisc}
\usepackage{xcolor}
\definecolor{orange}{RGB}{255,127,0}

\usepackage{hyperref}

\begin{document}

%\preprint{APS/123-QED}

\title{A strontium optical lattice clock with \num{1e-17} uncertainty and measurement of its absolute frequency}

\author{Richard~Hobson$^{1,*}$,
        William~Bowden$^{1, 2,*}$, 
        Alissa~Silva$^1$, 
        Charles~F.~A.~Baynham$^{1, 2}$, 
        Helen~S.~Margolis$^{1, 2}$, 
        Patrick~E.~G.~Baird$^2$,
        Patrick Gill$^{1, 2}$,
        Ian~R. Hill$^1$}
\address{$^1$National Physical Laboratory, Hampton Road, Teddington TW11 0LW, United Kingdom}
\address{$^2$Clarendon Laboratory, Parks Road, Oxford OX1 3PU, United Kingdom}
\address{$^*$Both authors contributed equally to this work.}
%\email{$^*$richard.hobson at npl.co.uk}

\date{\today}% It is always \today, today,

\begin{abstract}

We present a measurement of the absolute frequency of the 5s$^2$ $^1$S$_0$ to 5s5p $^3$P$_0$ transition in $^{87}$Sr, which is a secondary representation of the SI second. We describe the optical lattice clock apparatus used for the measurement, and we focus in detail on how its systematic frequency shifts are evaluated with a total fractional uncertainty of \num{1e-17}. Traceability to the International System of Units is provided via comparison to International Atomic Time (TAI). Gathering data over 5- and 15-day periods, with the lattice clock operating on average 74\% of the time, we measure the frequency of the transition to be \num{429228004229873.1}(5)~Hz, which corresponds to a fractional uncertainty of $1 \times 10^{-15}$. We describe in detail how this uncertainty arises from the intermediate steps linking the optical frequency standard, through our local time scale UTC(NPL), to an ensemble of primary and secondary frequency standards which steer TAI. The calculated absolute frequency of the transition is in good agreement with recent measurements carried out in other laboratories around the world.
\end{abstract}

\maketitle
\ioptwocol
%\tableofcontents

\section{Introduction}

%The precision of optical atomic clocks based on trapped ions \cite{Huntemann2016,Chou2010} and neutral atoms \cite{Nicholson2015,Nemitz2016} is making rapid progress, with systematic uncertainties now routinely surpassing those of the most precise Cs fountains \cite{Heavner2014,Li2011,Guena2012}. As well as the realization of stable time scales \cite{Grebing2016, Hachisu2018} and potentially supporting a future redefinition of the SI second \cite{Gill2011}, optical clocks have also been used as tools for geodesy \cite{Lisdat2015, Grotti2018}, relativity \cite{Delva2017,Chou2010a}, the search for dark matter \cite{Derevianko2014}, and the search for possible variations in fundamental constants \cite{Huntemann2014,Godun2014}.

The precision of optical atomic clocks based on trapped ions \cite{Huntemann2016,Brewer2019} and neutral atoms \cite{bothwell2019,Nemitz2016,McGrew2018b} is making rapid progress, with estimated systematic uncertainties now routinely lower than those of the most accurate Cs fountains \cite{Heavner2014,Li2011,Guena2012}. %As well as potentially supporting time scales \cite{Grebing2016, Hachisu2018, Yao2018} and a future redefinition of the SI second \cite{Gill2011, Riehle2018}, optical clocks are also promising tools for tests of relativity \cite{Delva2017,Chou2010a}, geodesy \cite{Lisdat2015, Grotti2018}, the search for dark matter \cite{Derevianko2014, wcislo2018}, and the search for possible variations in fundamental constants \cite{Huntemann2014,Godun2014}.
As well as potentially supporting time scales \cite{Grebing2016, Hachisu2018, Yao2019,Milner2019} and a future redefinition of the SI second \cite{Gill2011, Riehle2018}, optical clocks are also promising tools for geodesy \cite{Lisdat2015, Grotti2018} and for carrying out tests of fundamental physics including violations of relativity \cite{Delva2017,Chou2010a,Takamoto2020}, signatures of dark matter \cite{Derevianko2014, wcislo2018}, or time variation of fundamental constants \cite{Huntemann2014,Godun2014}.

One of the most successful approaches to atomic clock-making has been to use an optical lattice clock (OLC) configuration, in which a highly forbidden $^1$S$_0$ to $^3$P$_0$ transition is probed in an ensemble of laser-cooled neutral atoms trapped in a magic-wavelength optical lattice \cite{Ye2008}. OLCs are being pursued based on a wide range of alkaline earth-like atomic species, including Yb \cite{Yasuda2012, Hinkley2013, Kim2017, Pizzocaro2019}, Hg \cite{Yamanaka2015,Tyumenev2016}, Cd \cite{Kaneda2016}, and Mg \cite{Kulosa2015}. In this report we present results from an OLC based on $^{87}$Sr, contributing a new data point to a rich history of absolute \cite{Boyd2007b, Campbell2008, Hong2009, Baillard2007, LeTargat2013, Lodewyck2016, Falke2011b, Falke2014, Matsubara2009, Yamaguchi2011, Hachisu2016, Hachisu2017, Akamatsu2014, Tanabe2015, Lin2015} and relative \cite{Nemitz2016,Lisdat2015,Yamanaka2015,Ushijima2015, Koller2017} frequency measurements with the same atom.

This report is organized as follows. In section~\ref{sec:apparatus} we outline the optical lattice clock apparatus. In section~\ref{sec:systematics} we describe how the systematic frequency shifts and their uncertainties are evaluated. Finally, in section~\ref{sec:absolute_freq} we describe the absolute frequency measurement of the optical lattice clock against International Atomic Time (TAI).

\section{Experimental apparatus} \label{sec:apparatus}

A detailed overview of NPL's $^{87}$Sr optical lattice clock hardware has been given in previous work \cite{Hill2016,Donnellan2019}. Only some key parts of the experimental apparatus are recapitulated here. %There have been two minor changes to the hardware since the earlier work: a \SI{707}{\nano\meter} repump laser has been replaced by a different laser at \SI{497}{\nano\meter}, and the optical lattice beam waist has been increased from \SI{45}{\micro\meter} to \SI{65}{\micro\meter}.

The Sr atomic source consists of an oven heated to approximately \SI{800}{\kelvin}, followed by a \SI{30}{\centi\meter} transverse-field permanent-magnet Zeeman slower, details of which have previously been presented in \cite{Hill2014}. The atomic beam then enters the science chamber, a steel spherical octagon with radius \SI{112}{\milli\meter}, in which the atoms are captured and then cooled in a two-stage magneto-optical trap (MOT). The first \lq blue\rq\, MOT operates for between \SI{100}{\milli\second} and \SI{300}{\milli\second} on the 5s$^2$~$^1$S$_0$ to 5s5p~$^1$P$_1$ transition at \SI{461}{\nano\meter}. The large scatter rate of $2\pi\times \SI{30}{\mega\hertz}$ on this blue transition enables the efficient capture of atoms from the slowed atomic beam, but limits the MOT temperature to around \SI{2}{\milli\kelvin}. In order to prevent atoms from being shelved from the blue MOT into the 5s5p~$^3$P$_{0,2}$ states, repump lasers are applied at \SI{497}{\nano\meter} and \SI{679}{\nano\meter}, enhancing the blue MOT lifetime from around \SI{18}{\milli\second} to \SI{2}{\second}. For efficient repumping of $^{87}$Sr, transitions out of all five hyperfine $^3$P$_2$ states are addressed by tuning the \SI{497}{\nano\meter} laser to the $F = 11/2$ to $F^\prime = 11/2$ transition and modulating at \SI{710}{\mega\hertz} to a depth of around \SI{1.7}{\radian} using a waveguide electro-optic modulator (EOM). Further explanation of the repumping scheme used here, including a term diagram for \textsuperscript{87}Sr, can be found in Ref. \cite{Hobson2020}.

After the blue MOT, a \lq red\rq\, MOT is operated for a total of \SI{230}{\milli\second} on the 5s$^2$~$^1$S$_0$ to 5s5p~$^3$P$_1$ transition at \SI{689}{\nano\meter}. The low scatter rate of $2\pi\times \SI{7.5}{\kilo\hertz}$ on the red transition facilitates cooling to the low \si{\micro\kelvin} range in a two-stage sequence \cite{Katori1999b,Loftus2004}. An initial broadband red MOT stage lasting \SI{80}{\milli\second}, where the cooling laser is frequency modulated to cover a \SI{2}{\mega\hertz} spectrum, captures hot atoms from the blue MOT. Next, a single-frequency red MOT lasting \SI{160}{\milli\second} further cools the atoms to \SI{2}{\micro\kelvin} and compresses the cloud to a diameter of around \SI{200}{\micro\meter}, at which point the atoms automatically load into the co-located optical lattice trap. Throughout the red MOT, two separate laser frequencies are applied so that the $F = 9/2$ to $F^\prime = 9/2$ and the $F = 9/2$ to $F^\prime = 11/2$ transitions are both driven simultaneously---this \lq stirring\rq\, technique \cite{Mukaiyama2003} significantly improves the red MOT lifetime, enabling much more efficient loading of atoms into the lattice.

The \SI{813}{\nano\meter} optical lattice beam is oriented vertically and focused to a waist of \SI{65}{\micro\meter} at the atoms before being collimated and retro-reflected to form a 1D standing wave. We use a Ti:Sapphire laser as the \SI{813}{\nano\meter} source, which is frequency stabilized to a transfer cavity whose length is itself stabilized to the \SI{698}{\nano\meter} clock laser to ensure long-term stability. Before the beam is delivered to the atoms, it propagates through a volume holographic grating optical bandpass filter with \SI{12}{\giga\hertz} full width half maximum, which strongly suppresses any spectral impurities arising from laser-amplified spontaneous emission or any other sources. The system can supply up to \SI{1}{\watt} at the atoms, generating lattice trap depths up to $U_0$~$=$~$200\,E_\mathrm{r}$, where $E_\mathrm{r} = h\times\SI{3.4}{\kilo\hertz}$ is the lattice photon recoil energy.

Before commencing clock spectroscopy, three further stages of state preparation are implemented on the lattice-trapped atoms. First, the atoms are optically pumped into either the 5s$^2$\,$^1$S$_0$~$M_F = 9/2$ or the 5s$^2$\,$^1$S$_0$~$M_F = - 9/2$ state using a pulse of $\sigma^+$ or $\sigma^-$ polarized light at \SI{689}{\nano\meter} addressing the 5s$^2$\,$^1$S$_0$~$F = 9/2$ to 5s5p\,$^3$P$_1$~$F^\prime = 9/2$ transition. The optical pumping typically achieves a spin-polarization efficiency of 75\,\%. Second, a \lq spilling\rq\, stage is used to select a colder atomic sample at around \SI{2}{\micro\kelvin}: the lattice depth is linearly ramped down to $26\,E_\mathrm{r}$ in \SI{20}{\milli\second}, held for \SI{20}{\milli\second} to let the hot atoms escape, and then ramped back up to the operating depth of $52\,E_\mathrm{r}$ for clock spectroscopy. Third, a \SI{22}{\milli\second} Rabi $\pi$ pulse is implemented at \SI{698}{\nano\meter} in a quantization field of \SI{72}{\micro\tesla}, resonant with the 5s$^2$\,$^1$S$_0$ $M_F = \pm 9/2$ to 5s5p\,$^3$P$_0$ $M^\prime_F = \pm 9/2$ transition and therefore not resonant with transitions from other Zeeman sublevels. The atoms remaining in the ground state are then pushed out of the lattice using a pulse of \SI{461}{\nano\meter} light, leaving a sample of cold atoms of high purity in the excited $M^\prime_F = 9/2$ or $M^\prime_F = - 9/2$ state.

To realize the frequency standard, the clock transition is interrogated with a \SI{200}{\milli\second} Rabi $\pi$ pulse using a \SI{698}{\nano\meter} clock laser, and the resulting excitation fraction is used to steer the laser toward the atomic resonance. The clock laser consists of a home-built extended-cavity diode laser (ECDL) which is pre-stabilized with \SI{1.5}{\mega\hertz} bandwidth to a replica of the \lq football\rq\, cavity in Ref \cite{Ludlow2007}. This \SI{698}{\nano\meter} laser is then phase-locked through an acousto-optic modulator to another, more stable laser operating either at \SI{934}{\nano\meter} or \SI{1064}{\nano\meter}. In either case the flicker-noise fractional frequency instability transferred to \SI{698}{\nano\meter} is around \num{5e-16}. The laser phases are compared over a self-referenced optical frequency comb in a similar arrangement as in Ref \cite{Hagemann2013}, exploiting a transfer oscillator scheme \cite{Telle2001} to cancel noise in the comb repetition rate.

\section{Systematic frequency shifts}
\label{sec:systematics}

There are several systematic frequency shifts influencing optical lattice clocks, all of which must be characterized and corrected for in order to realize an accurate frequency standard. Here we describe how these shifts are characterized in our lattice clock system, reaching a total systematic uncertainty of \num{1.0e-17}. The individual contributions to the uncertainty budget are outlined in table \ref{tab:systematics}.

\begin{table}
\caption{\label{tab:systematics}Uncertainty budget for the NPL Sr lattice clock. Reported uncertainties correspond to 68\% confidence intervals. All values are in units of  $1\times10^{-18}$. }

\begin{tabular}{lcc}
\toprule
Systematic effect & Correction & Uncertainty \\
\hline
%!!!There are no footnotes in the compiled document.
BBR chamber\tablefootnote{The BBR correction and uncertainty change with time---see text} & 4875.1 & 7.0 \\
BBR oven & 0.5 & 0.5\\
Quadratic Zeeman & 287.0 & 3.0\\
Lattice & 0.5 & 4.4\\
Collisions &  0.9  &  3.8 \\
Background gas & 2.0 & 2.0\\
DC Stark & 0.016 & 0.016\\
Probe Stark & 1.0 & 0.4\\
%AOM phase chirp & 0 & $<0.1$\\
Servo Error\tablefootnote{The servo error is below \num{2e-18} for datasets longer than 1 hour---see text} & 0 & 2\\
\hline
\textbf{Total Correction} & 5167 & 10\\
\bottomrule
\end{tabular}

\end{table}

\subsection{Blackbody radiation shift}

The largest systematic shift in the Sr optical lattice clock is from the blackbody radiation (BBR) emitted by the room-temperature vacuum chamber. Following Ref \cite{Safronova2013}, the BBR-induced fractional frequency shift can be split into static and dynamic components:

\begin{align}
y_\mathrm{BBR} = \beta_\mathrm{st} \left(\frac{T}{T_0}\right)^4\Bigg[1 + \eta_1\left(\frac{T}{T_0}\right)^2 &+\eta_2\left(\frac{T}{T_0}\right)^4 \nonumber \\ &+ \eta_3\left(\frac{T}{T_0}\right)^{6}\Bigg] \label{eq:BBR}
\end{align}
where we have chosen an arbitrary reference temperature $T_0 = \SI{300}{\kelvin}$. Higher-order dynamic corrections are below \num{1e-18}, while magnetic dipole and electric quadrupole interactions with the BBR field contribute only at the \num{6e-20} level and can therefore be neglected \cite{Porsev2006}.

The static coefficient in Equation \ref{eq:BBR} can be calculated from the DC polarizability measurement in Ref \cite{Middelman2012} as $\beta_\mathrm{st} = -4962.93(14) \times 10^{-18}$. The dynamic corrections can be calculated from various line strength data, with the largest contribution being from the \SI{2.6}{\micro\meter} 5s5p~$^3$P$_0$ to 5s4d~$^3$D$_1$ transition measured in Ref \cite{Nicholson2015}, giving $\beta_\mathrm{st}\eta_1 = -300.7(14)\times 10^{-18}$, $\beta_\mathrm{st}\eta_2 = -37.6(2)\times 10^{-18}$ and $\beta_\mathrm{st}\eta_3 = -7.97(3)\times 10^{-18}$. (Note that the uncertainties in these dynamic coefficients are strongly correlated, so we calculate a total uncertainty by summing linearly rather than in quadrature). At our operating temperature of \SI{294}{\kelvin}, the total systematic uncertainty introduced due to imperfect knowledge of the BBR coefficients is \num{1.4e-18}.

However, the main limitation in our lattice clock system is not from theory, but rather from the experimental uncertainty in the BBR temperature $T/T_0$. To measure this temperature, eleven Pt100 sensors are affixed to the outside of the vacuum chamber in a range of positions. Each sensor is calibrated with an uncertainty of \SI{10}{\milli\kelvin} in accordance with the principles outlined in reference \cite{Preston1990}. We choose to locate seven of the sensors directly on the viewports and only four on the steel chamber because the emissivity of fused silica is much greater than that of polished steel. The dominant source of BBR uncertainty is not from the sensor calibration, but instead from temperature differences between different parts of the chamber.

To account for the range of temperature readings we follow the same approach as in Ref \cite{Falke2014}, which models the representative BBR temperature as a rectangular probability distribution between the highest sensor reading $T_\mathrm{max}$ and the lowest $T_\mathrm{min}$. This is in accordance with BIPM’s ‘GUM: Guide to the Expression of Uncertainty in Measurement’ which recommends adopting such a distribution when only the bounds on a quantity are known \cite{bipm2008}. This yields an estimated temperature of $(T_\mathrm{max} + T_\mathrm{min})/2$ with an uncertainty of $(T_\mathrm{max} - T_\mathrm{min})/\sqrt{12}$.

%To account for the range of temperature readings we follow the same approach as in Ref \cite{Falke2014}: the BBR temperature is modelled as following a rectangular probability distribution between the highest sensor reading $T_\mathrm{max}$ and the lowest $T_\mathrm{min}$, yielding an estimated temperature of $(T_\mathrm{max} + T_\mathrm{min})/2$ with an uncertainty of $(T_\mathrm{max} - T_\mathrm{min})/\sqrt{12}$.

\begin{figure}[ht]
    \centering
    \includegraphics[width=0.48\textwidth]{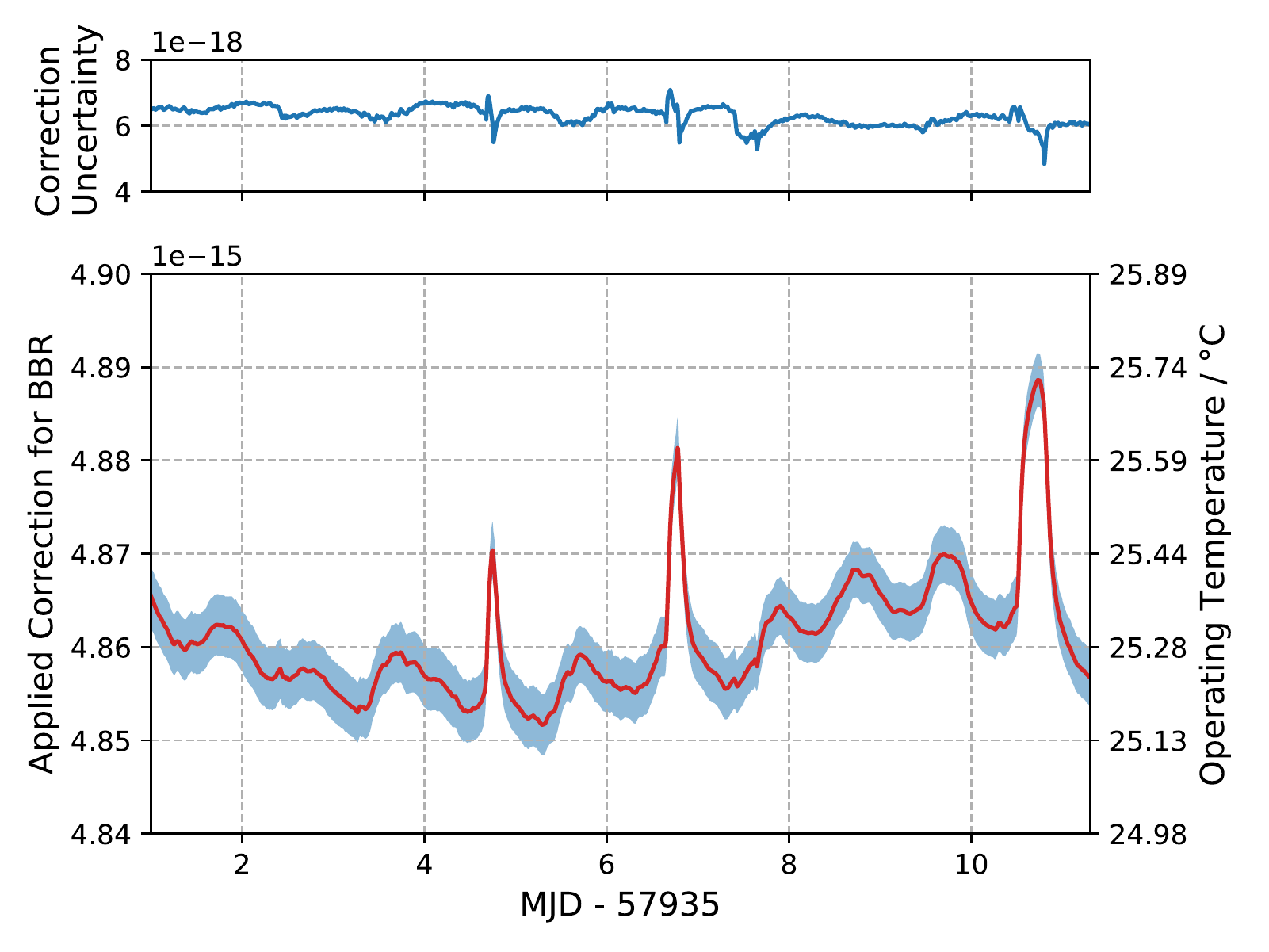}
    \caption{The lower plot shows the fractional frequency BBR correction over the first 10 days of the campaign along with the corresponding estimate of the equivalent operating temperature. The blue shaded region represents the 1-sigma uncertainty which is also plotted in the upper plot for clarity. This uncertainty is primarily set by the maximum gradient across the chamber. \label{fig:BBR}}
\end{figure}

In a typical experimental run the temperature of the chamber slowly drifts over time, mostly due to fluctuations in the temperature of the water used to cool the MOT coils mounted to the side of the chamber. Therefore, to minimize the uncertainty, we continuously monitor and log the temperature sensor readings on a one-second grid and calculate time-dependent corrections as shown in figure \ref{fig:BBR}. The peak-to-peak temperature gradient across the chamber is usually around \SI{350}{\milli\kelvin}, yielding a BBR uncertainty of around \num{7e-18}. %\textcolor{blue}{\st{However, the temperature of the cooling water became less stable at some points during the absolute frequency measurement in this report, causing temperature gradients of up to 1 K and a resulting frequency uncertainty up to 2e-17}}.

As a consistency check we also developed an alternative model for the total BBR field using a solid-angle- and emissivity-weighted sum of contributions from the various components of the science chamber, similar to the approach presented in Ref \cite{BloomThesis}. In a typical experimental run the two models agree with each other to within much less than their uncertainties. However, we conservatively base our final estimate on the rectangular distribution model since it returns a larger uncertainty and doesn't rely on assumptions regarding material properties and geometries unlike the emissivity-weighted, solid-angle approach.

Finally we must consider the contribution to the BBR environment from the Sr oven source, which is heated to around \SI{800}{\kelvin} and has a direct line of sight to the atoms. We model this effect using a similar approach as for the leakage into the cryogenic BBR enclosure in Ref \cite{Middelmann2011}, but in our system we find a negligible oven contribution on the order of \num{2e-19}---the BBR leakage into the main chamber is strongly suppressed by the presence of two $\SI{1}{\milli\meter}$-diameter apertures along the atomic beam, both placed before the Zeeman slower at a distance of around \SI{0.5}{\meter} from the atoms with a separation of $\SI{0.1}{\meter}$.

\subsection{Zeeman shift}

The clock transition is interrogated in a bias field of \SI{72}{\micro\tesla} in order to ensure that the different Zeeman transitions are well resolved from each other. With the bias field applied, the $M_F = 9/2$ to $M^\prime_F = 9/2$ transition has a linear sensitivity to fluctuations in the magnetic field strength of $1.139(5) \times 10^{-14}\,\si{\per\micro\tesla}$ in fractional frequency units \cite{Baillard2008,Boyd2007a}.

In order to provide immunity from the linear Zeeman shift, two clock servos are combined with spin-polarized samples of atoms in the $M_F = +9/2$ and $M_F = -9/2$ stretched states respectively \cite{Takamoto2006}. The clock servo software then logs the average frequency $\nu_\mathrm{avg} =  \left(\nu_{+9/2}+\nu_{-9/2}\right)/2$, for which the linear Zeeman shift cancels out. In theory, this cancellation could be compromised by drift in the bias magnetic field between consecutive servo cycles; however, we observe a sufficiently small magnetic field drift of \SI{20}{\nano\tesla} over \SI{e5}{\second} to imply a negligible effect from drifts in the linear Zeeman shift.

By contrast, the quadratic Zeeman shift is not cancelled by averaging the two stretched transitions. To characterize the quadratic shift, the frequency splitting $\nu_{+9/2}-\nu_{-9/2}$ is continuously logged and the shift coefficient of $-5.75(4)\times 10^{-16}\,\si{\per\kilo\hertz\squared}$ \cite{Nicholson2015} is applied. For the splitting of \SI{706}{\hertz} used during the measurement in this report, we calculate a systematic shift of $-2.87(3) \times 10^{-16}$.

\subsection{Lattice Stark shift}

The optical lattice trap is operated close to the magic wavelength where the shifts of the excited and ground states should be the same \cite{Ye2008}. Nonetheless, there are several systematic shifts caused by the lattice trapping field, for example: residual electric dipole (E1) shifts due to deviations from the magic wavelength; magnetic dipole (M1) and electric quadrupole (E2) interactions; and fourth-order perturbations from the electric dipole interaction (also known as hyperpolarizability). Since the different interaction terms have different spatial profiles in the 1D lattice, a complete treatment must take into account details of the atomic motion within the lattice potential \cite{Brown2017}.

In this report we apply a simplified model of the lattice shift, neglecting some of the higher-order motional corrections which only contribute below the \num{2e-18} level in our operating conditions. In the simplified model, the total fractional frequency shift can be written:

\begin{align}
    y_L = \Delta\alpha^\prime_\mathrm{E1} \left(\frac{U_\mathrm{eff}}{E_r}\right) &+ \beta^\prime \left(\frac{U_\mathrm{eff}}{E_r}\right)^2 \nonumber \\ &+ \Delta\alpha^\prime_\mathrm{E2M1}\left(n_z+\frac{1}{2}\right) \left(\frac{U_0}{E_r}\right)^{\frac{1}{2}} \label{eq:lattice_shift}
\end{align}
where $n_z$ is the axial motional state, and we define an effective trap depth $U_\mathrm{eff} = U_0 - kT_r$ to take into account that the finite radial temperature $T_r$ reduces the mean lattice intensity seen by the atoms---in a classical approximation, using the equipartition theorem, the atoms possess a potential energy of $kT_r/2$ along each of the two radial degrees of freedom. The axial temperature $T_z \approx \SI{1.5}{\micro\kelvin}$ is not included when calculating the effective trap depth since along the lattice axis the atoms are in the non-classical regime $\left<n_z\right> \approx 0.25 \ll 1$.

\begin{figure}[t]
    \centering
    \includegraphics[width=0.48\textwidth]{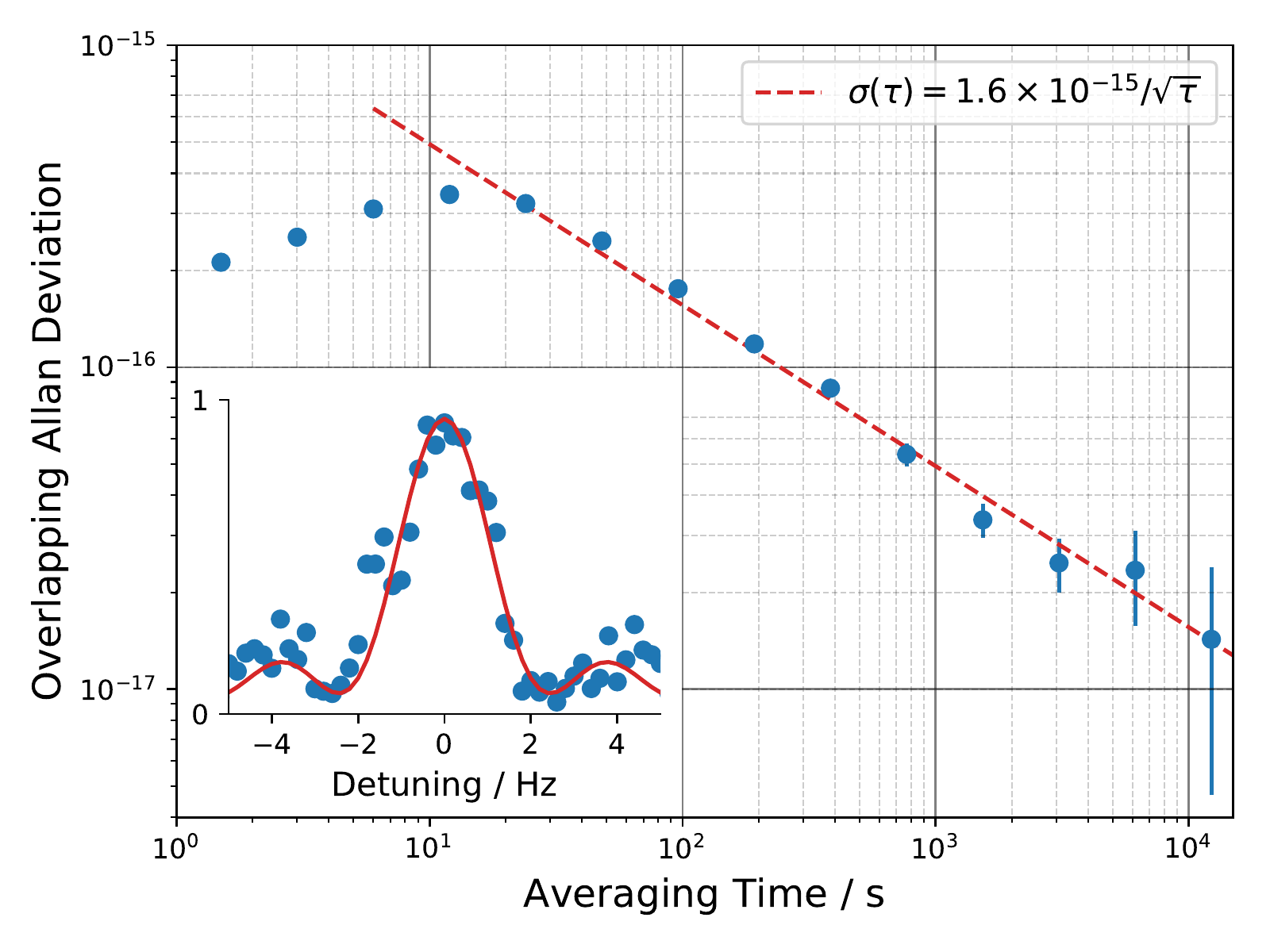}
    \caption{ Overlapping Allan deviation of the frequency difference between two interleaved atomic servos operating at different lattice depths over a continuous \SI{14}{\hour} period. By measuring the offsets between the servos we can estimate the lattice induced Stark shift under normal operating conditions. For our system the interleaved instability, shown in red, is $1.6\times 10^{-15}/\sqrt{\tau}$. Inset shows the line shape for a \SI{300}{\milli\second} Rabi pulse which has a Fourier limited linewidth of \SI{2.7}{\hertz}.\label{fig:rabi}}
\end{figure}

To evaluate Equation \ref{eq:lattice_shift}, we apply the E2M1 coefficient $\Delta\alpha^\prime_\mathrm{E2M1}$~$=$~$0.0(72)$~$\times$~$10^{-19}$ and hyperpolarizability coefficient $\beta^\prime$~$=$~$1.07(42)$~$\times$~$10^{-21}$, as measured in Ref. \cite{Westergaard2011}. The residual E1 coefficient $\Delta\alpha^\prime_\mathrm{E1}$, which depends strongly on local operating conditions such as the lattice wavelength, is extracted by fitting to frequency shift data between interleaved servos at different lattice depths ranging between $52\,E_\mathrm{r}$ and $156\,E_\mathrm{r}$. The frequency instability of one of these interleaved datasets is shown in figure \ref{fig:rabi}. The fitting script is run multiple times in a Monte Carlo simulation using randomly generated E2M1 and hyperpolarizability coefficients, so the uncertainties in these higher-order shifts are propagated appropriately. Motional sideband scans are implemented at each operating depth in order to evaluate $U_0$, $T_z$ and $T_r$ \cite{Blatt2009}, and the lattice wavelength and lattice polarization (parallel with the magnetic field) are both kept constant throughout all measurements. Combining several days of interleaved lattice shift data, we reach a total shift of $-0.5(44) \times 10^{-18}$ at the operating depth of $52\,E_\mathrm{r}$.%, with the largest uncertainty arising from statistics of the interleaved frequency difference data.

As noted in Ref \cite{Nicholson2015}, the lattice shift measurement can be distorted by parasitic collisional effects: a deeper lattice will compress the atoms into a smaller volume, causing a differential collisional shift that could be mistaken as a lattice shift. Furthermore, loading atoms into a deeper trap or ramping the trap depth after loading can change the atomic temperature---in turn changing the collisional shift \cite{Brown2017}. To estimate how this affects the measured lattice shift, we run the Monte Carlo simulation using two different models for the scaling of atom density $\rho(U)$ with trap depth:  $ \rho\propto U_0^{3/2}$, which assumes $T_z$ and $T_r$ are constant with trap depth, and $\rho\propto U_0^{3/4}$, which assumes the mean thermal occupancy number is constant. The latter closely resembles our experimental conditions, since we load atoms into the lattice at a fixed trap depth before adiabatically ramping to the final set point used during spectroscopy. Based on data from our independent collisional shift evaluation in section \ref{sec:other_systematics}, and by operating with a reduced atom number at higher lattice set points, we estimate that the uncertain relationship between the collisional shift and the trap depth introduces an uncertainty in the estimated lattice shift of \num{1e-18}. 

%\In principle the lattice shift measurement could be distorted by parasitic collisional effects---a deeper lattice will compress the atoms into a smaller volume, potentially causing a differential collisional shift masquerading as a lattice shift. We suppress the possible parasitic effect of collisions using the method in Ref \cite{Nicholson2015}: during lattice shift evaluation the atom number is adjusted as $N \propto U_0^{-3/2}$ to roughly compensate the collisional shift, which would scale as $U_0^{3/2}$ \textcolor{blue}{under the assumption that atomic temperature remains constant with trap depth}. Using data from our independent collisional shift evaluation in section \ref{sec:other_systematics}, \textcolor{blue}{and including in the model an uncertainty in the shift scaling in the range $U_0^{3/4}$ to $U_0^{3/2}$ depending on whether atoms maintain constant $T$ or constant $\left<n\right>$ as $U_0$ is varied}, we estimate this suppression technique to be effective with a residual uncertainty below \num{5e-19}.

\subsection{Other systematic shifts}\label{sec:other_systematics}

\textit{Collisions:} The cooling sequence described in section \ref{sec:apparatus} prepares high purity samples of atoms in an identical internal state, implying that $s$-wave collisions should be strongly suppressed by the Fermi exclusion principle. Nonetheless we search for any small residual $s$-wave or $p$-wave collisional shifts by measuring the frequency difference between interleaved servos containing atom numbers $N_0$ and $4.2N_0$ respectively. The observed frequency difference of $-3(12)\times 10^{-18}$ between these servos implies a systematic shift at our operating atom number $N_0$ of  $-0.9(38) \times 10^{-18}$.

\textit{Background gas collisions:} In our science chamber the vacuum-limited lifetime both for magnetically-trapped 5s5p\,$^3$P$_2$ atoms and for lattice-trapped 5s$^2$\,$^1$S$_0$ atoms is observed to be \SI{8}{\second}. Applying the model in Ref. \cite{Gibble2013}, and assuming a background dominated by hydrogen (or by other gases with similar $C_6$ coefficients \cite{Mitroy2010a}), the associated collisional frequency shift is estimated as $-2\times 10^{-18}$. Since this model of the background gas shift is yet to be experimentally verified, the uncertainty in the shift is also taken to be $2\times 10^{-18}$.

\textit{DC Stark shift:} We evaluate the background electric field using spectroscopy on a Rydberg transition to 5s75d\,$^1$D$_2$, as previously described in \cite{Bowden2017}. The large steel chamber with comparatively small viewports proves to be quite an effective shield against background electric fields, yielding a systematic shift of $-1.6(16) \times 10^{-20}$.

%\textit{Probe Stark shift:} We estimate the probe Stark shift using the experimentally measured natural lifetime of the 5s5p\,$^3$P$_0$ state of $330(140)$~s \cite{Dorscher2018}. If we include the Clebsch-Gordan coefficient 0.9045 and the Lamb-Dicke parameter $1-\eta_z^2 = 0.90$, both of which slightly increase the intensity required to drive a \SI{200}{\milli\second} Rabi $\pi$ pulse on the clock transition, we estimate a probe Stark shift of $-1.0(4) \times 10^{-18}$.

 \textit{Probe Stark shift:} We estimate the probe beam intensity required to drive the \SI{200}{\milli\second} Rabi $\pi$ pulse on the clock transition using the experimentally measured natural lifetime of the 5s5p\,$^3$P$_0$ state of $330(140)$~s \cite{Dorscher2018}. We also include the Clebsch-Gordan coefficient 0.9045 and the Lamb-Dicke parameter $1-\eta_z^2 = 0.90$, both of which slightly increase the required intensity. This estimated intensity, when multiplied by the theoretical differential polarisability at \SI{698}{\nano\meter} between the two clock states \cite{hill2012}, results in an estimated probe-induced Stark shift of $-1.0(4) \times 10^{-18}$.

\textit{Doppler shifts:} To a good approximation, the position of each atom is defined by the position of the lattice site in which it is trapped. Since the lattice wavelength is well controlled, the motion of the lattice sites is mostly set by the motion of the lattice retro-reflecting mirror. To avoid first-order Doppler shifts, it is therefore important to make sure that the clock probe beam phase is as stable as possible relative to the lattice retro-reflector. In our system, we actively phase-stabilize the delivered clock probe light \cite{Ma1994} at a reference sampler placed close to the lattice retro-reflector, leaving only a short uncompensated path of \SI{20}{\centi\meter} in free space. The beat signal used in this phase stabilization loop is logged on a frequency counter throughout any measurement so that any cycle slips can be detected and the corresponding data points discarded (though normally no such glitches are detected). %{\color{blue} \st{There is no a priori reason to believe that the short uncompensated path would result in a significant systematic shift at the 1e-18 level: sources of vibrations are kept to a minimum, and all acoustic resonances are high-frequency compared with the 200 ms Rabi pulse time so that the effect of residual vibrations should be averaged out. Therefore the systematic shift from the uncompensated path is probably negligible and is not included in table \ref{tab:systematics}. However, to verify this assumption we interleave servos with different delays between the cooling stages and the clock probe pulse and could not resolve a shift at the level of our total systematic uncertainty.}}
The second-order Doppler shift from thermal motion at the atomic temperature of \SI{2}{\micro\kelvin} is below \num{e-20} and is therefore also omitted from table \ref{tab:systematics}.

%\textcolor{red}{\st{In order to evaluate possible residual Doppler shifts arising from the uncompensated path, e.g. from vibrations induced by mechanical shutters switched during the cooling sequence, we interleave servos with different delays between the cooling stages and the clock probe pulse and measure a frequency difference of $1 \pm 6 \times 10^{-18}$. Since the measured shift is consistent with zero with an uncertainty lower than other systematic shifts, and since there is no a priori reason to believe that a shift is likely to be present at the \num{e-18} level, we conclude that the Doppler shift is probably negligible and do not include it in table \ref{tab:systematics}. The second-order Doppler shift from thermal motion at the atomic temperature of \SI{2}{\micro\kelvin} is below \num{e-20} and is therefore also neglected.}}

\textit{AOM phase chirp:} In order to suppress the heating-induced phase chirp from the final \lq switching\rq\, acousto-optic modulator (AOM), the reference sampler for phase stabilization is placed after the AOM, retro-reflecting the $0^\mathrm{th}$-order beam. The idea behind this arrangement is that the $0^\mathrm{th}$-order beam traverses a very similar path through the AOM as the $1^\mathrm{st}$-order beam which probes the atoms. The effectiveness of this compensation technique has been verified by turning off the phase stabilization servo and measuring the phase chirp during the probe pulse at high RF drive power against a separately delivered phase-stabilized beam. With the servo re-engaged, we observe that the phase stabilization loop suppresses the AOM phase chirp by at least a factor of 5 (with this factor limited by statistics in the measurement). Combined with the reduced RF drive power used during clock operation, which results in an experimentally verified proportionate reduction in the phase chirp, we estimate a systematic shift from AOM phase chirp of less than \num{3e-22}.

\textit{Line pulling:} The nearest $\pi$-polarized transition is $M_F = 7/2$ to $M_F^\prime = 7/2$, which is split from the main clock transition by \SI{67}{\hertz}. When scanning the clock laser over the expected $M_F = 7/2$ transition we observe no excitation at the 3\% level, while on the main clock transition a contrast of 95\% is observed with a Fourier-limited line-width of \SI{4}{\hertz}. Combining these observations we calculate a maximum line-pulling effect of \num{5e-20}. If the probe beam were not perfectly $\pi$-polarized, then it would also be possible to observe line-pulling effects from the $M_F = 9/2$ to $M_F = 7/2$ transition. However, since we observe no excitation of this transition at the 3\% level when scanning over it, and since it should be detuned by \SI{210}{\hertz} from the main clock transition \cite{Boyd2007a}, the expected line pulling effect is well below \num{e-20}.

%\textit{Servo Error:} The clock servo is always trying to make sure that the local oscillator is steered exactly to atomic resonance, but lock offsets can emerge from two sources: (i) finite servo gain, which can allow the local oscillator to sag above or below the clock transition frequency (ii) asymmetric sampling of the local oscillator drift if left uncompensated between AOM updates. In order to mitigate the second problem, the local oscillator is de-drifted with AOM updates every \SI{100}{\milli\second}, leading to a maximum sawtooth frequency deviation of less than \SI{12}{\milli\hertz} given the cavity drift rate of \SI{140}{\milli\hertz\per\second}. THIS IS A BIT RUBBISH! Finally, to characterize the error due to finite servo gain, the atomic excitation is recorded and post-processed to recalculate the error signal and therefore the mean frequency offset. For datasets longer than \SI{1}{\hour} the servo offset is consistently below \num{2e-18}.

\textit{Servo Error:} The clock servo acts to ensure that the local oscillator is steered exactly to atomic resonance, but lock offsets can remain due to finite servo gain which allow the local oscillator to sag above or below the clock transition frequency. To characterize the error due to finite servo gain, the atomic excitation is recorded and post-processed to recalculate the error signal and therefore the mean frequency offset. We observe that on average the servo error scales as the inverse of the measurement time, and for datasets longer than \SI{1}{\hour} the servo offset is consistently below \num{2e-18}.

\section{Absolute frequency measurement} \label{sec:absolute_freq}

Determination of the absolute frequency of the strontium optical lattice clock requires measurement against a realization of the SI second. This can be achieved either via direct comparison to a local caesium primary standard, or indirectly by measuring the frequency versus International Atomic Time (TAI), which also provides traceability to the SI second. Historically, the majority of absolute frequency measurements have followed the direct approach, but more recently several groups reported values which were derived via comparison to TAI \cite{Park2013, Kim2017, Huang2016, Dube2017}, in some cases attaining uncertainties below \num{5e-16} \cite{Hachisu2017, Baynham2017} or even surpassing the best direct measurements using local primary frequency standards \cite{McGrew2018}. For the measurement in this report we adopted the latter approach, accessing the SI second through comparison against TAI.

\begin{figure*}[t]
    \centering
    \includegraphics[width=0.85\textwidth]{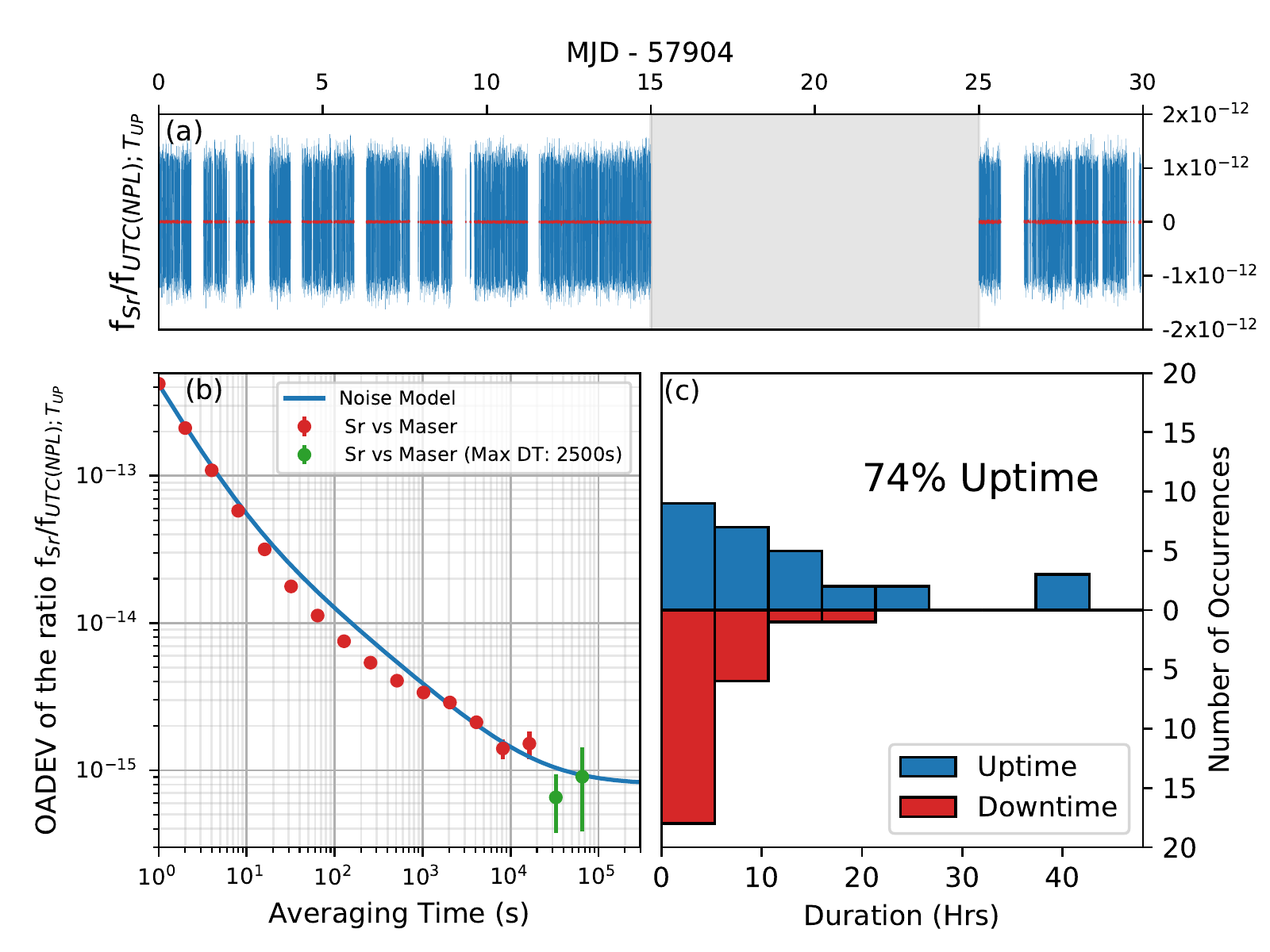}
     \caption{Figure (a) shows the ratio between the frequency of the optical clock transition in Sr and the frequency of the hydrogen maser as measured over the two campaign periods. The red curve represents a moving average over a 100 second window. Figure (b) shows the fractional frequency instability of the ratio between the OLC and the hydrogen maser averaged over the entire campaign as measured by the overlapping Allan deviation. The red Allan deviation points are based only on continuous datasets while green points are based on quasi-continuous datasets which can have downtime periods of less than 2500 seconds (see text for more details). From this instability data, we generate the stochastic noise model (blue curve) used in the Monte Carlo modelling of the uncertainty arising from downtime in the measurement. (c) Distribution of all continuous periods of uptime and downtime exceeding 10 minutes in length. The majority of downtime results from the unlocking of one of the lasers needed for cooling and trapping the Sr atoms. For most of these lasers, we have automatic recovery systems that can reacquire lock quickly without human intervention. As a result of this automation, the median downtime period over the entire campaign is 20 seconds. \label{fig:maser}}
\end{figure*}

The absolute frequency measurement is performed in several stages. First, the optical frequency must be compared to the scale interval of our local time scale UTC(NPL). Secondly, the local time scale is compared against TAI. This latter comparison is complicated by the fact that TAI is a virtual time scale that is computed on a monthly basis by the International Bureau of Weights and Measures (BIPM), as an intermediate step in the computation of Coordinated Universal Time (UTC). The time offset of UTC(NPL) from UTC is published at 5-day intervals in section 1 of BIPM's monthly Circular T bulletin. This means that it is only possible to calculate the frequency offset of the local time scale from TAI averaged over 5-day intervals. (Note that the scale interval of UTC is identical to that of TAI, since the two differ only by an integer number of leap seconds.) Finally, a correction must be applied to account for the deviation \textit{d} of the scale interval of TAI from the  SI second, published only as a monthly average in section 3 of Circular T. If the period of an optical frequency measurement does not coincide with these monthly reporting periods, it is also possible to request a custom computation of \textit{d} for specific periods corresponding to one or more 5-day Circular T reporting intervals \cite{Petit2018}. The procedure linking the local optical frequency reference to the SI second can be summarized by the following expression:

\begin{equation}\label{eq:ratioEquation}
\begin{split}
    \frac{f_{\mathrm{Sr}}}{f_{\mathrm{SI}}} =  \frac{f_{\mathrm{Sr}}}{f_{\mathrm{UTC(NPL)}}} \times 
 \frac{f_{\mathrm{UTC(NPL)}}}{f_{\mathrm{TAI}}} \times
    \frac{f_{\mathrm{TAI}}}{f_{\mathrm{SI}}}
\end{split}
\end{equation}

\noindent where $f(\mathrm{SI}) = $ \SI{1}{\hertz} by definition. In practice, the combination of optical clock downtime and the frequency instability of the local time scale complicates the measurement. This downtime can give rise to measurement error as the mean frequency of the local time scale during the time period when the optical clock was operational ($T_{\mathrm{UP}}$) may differ from its average over the total measurement period ($T_{\mathrm{ALL}}$). To account for this effect, we modify equation \ref{eq:ratioEquation} as follows: 
\begin{equation}\label{eq:ratioEquationMod}
\begin{split}
    \frac{f_{\mathrm{Sr}}}{f_{\mathrm{SI}}} = & \frac{f_{\mathrm{Sr}}}{f_{\mathrm{UTC(NPL)};{\mathrm{T}_{\mathrm{UP}}}}} \times 
    \frac{f_{\mathrm{UTC(NPL)};{\mathrm{T}_{\mathrm{UP}}}}}{f_{\mathrm{UTC(NPL)};\mathrm{T}_{\mathrm{ALL}}}} \times\\ &\frac{f_{\mathrm{UTC(NPL)};\mathrm{T}_{\mathrm{ALL}}}}{f_{\mathrm{TAI};\mathrm{T}_{\mathrm{ALL}}}} \times
    \frac{f_{\mathrm{TAI};\mathrm{T}_{\mathrm{ALL}}}}{f_{\mathrm{SI}}}.
\end{split}
\end{equation}
%where $\mathrm{T}_{\mathrm{UP}}$ is the time during which the clock was operational.

\noindent Using equation \ref{eq:ratioEquationMod}, we determine the absolute frequency of the 5s$^2$ $^1$S$_0$ to 5s5p $^3$P$_0$ transition in $^{87}$Sr over two separate periods during the month of June 2017 %\textcolor{blue}{\st{(MJD 57904-57919 and MJD 57929-57934)} 
(MJD 57904-57918 and MJD 57929-57933)  as shown in figure \ref{fig:maser}. Below we describe how a value for each ratio in the above expression is determined for these two measurement intervals. 

\subsection{Local frequency ratio: $f_{\mathrm{Sr}}/f_{\mathrm{UTC(NPL)};\mathrm{T}_{\mathrm{UP}}}$ }

The local frequency ratio $f_{\mathrm{Sr}}/f_{\mathrm{UTC(NPL)};\mathrm{T}_{\mathrm{UP}}}$ is determined by using a femtosecond optical frequency comb referenced to a \SI{10}{\mega\hertz} signal produced by a hydrogen maser. The same maser is used to generate a one pulse-per-second signal from its 10~MHz output which serves as the basis of UTC(NPL), hereby providing a direct link between the optical frequency and the local time scale. The frequency comb has been verified to introduce negligible uncertainty in such optical-microwave frequency comparisons. The RF beat signals are $\pi$-counted, with a one second gate time, using K+K FXE frequency counters.

The main source of uncertainty in this measurement comes from the distribution of the 10~MHz maser signal to the frequency comb laboratory, and the subsequent  synthesis in that laboratory of an 8 GHz signal against which the repetition rate of the frequency comb is measured. Potential time-varying phase shifts are monitored by dividing the 8~GHz signal by 800 and comparing the resulting 10~MHz signal with the original signal from the hydrogen maser using a phase comparator. Based on this round-trip data, we estimate the RF~distribution and synthesis to contribute an uncertainty of \num{1e-16} to the frequency ratio measurement. %\textcolor{blue}{\st{We also have to account for uncertainty introduced by the long-term frequency instability of the optical frequency standard. Based on lattice Stark shift evaluations carried out at seven points distributed throughout the measurement period, and based on our models for the blackbody radiation shift and other systematic effects, we estimate the long-term instability of the optical frequency standard to be below $1\times10^{-17}$. Further, we have directly verified the instability to be below $2\times10^{-17}$ at 1~day based on a comparison against similar frequency standards via an optical fibre network.} }

We also account for uncertainty introduced by the frequency instability of the optical frequency standard. Extrapolating the estimated white frequency noise as $2\times10^{-15}\,\tau^{-1/2}$, a conservative upper bound based on measurements against another OLC over an optical fibre network \cite{Delva2017}, for the total measurement uptime, we estimate the statistical uncertainty arising from frequency instability of the OLC to be \num{2e-18} and \num{3.5e-18} for the two measurement periods.

\subsection{Downtime correction: $f_{\mathrm{UTC(NPL);T_\mathrm{UP}}}/f_{\mathrm{UTC(NPL);T_\mathrm{ALL}}}$}

%\subsection{Estimation of the downtime correction}

To estimate the correction and uncertainty associated with the ratio of  $f_{\mathrm{UTC(NPL);T_\mathrm{UP}}}$ to $f_{\mathrm{UTC(NPL);T_\mathrm{ALL}}}$ arising from measurement downtime, we follow a similar approach to \cite{Baynham2017, Hachisu2015, yu2007}. The frequency instability of the maser can be split into deterministic (e.g. linear drift) and stochastic (e.g. white and pink frequency noise) parts. Given the predictable nature of the deterministic part, the resulting offset can be directly computed. For example, for the linear drift exhibited by the hydrogen maser, the frequency offset is the drift rate multiplied by the difference in time between the centre of the measurement window and the average time of the periods in which the clock was operational. Following this procedure, we estimate the deterministic downtime correction based on the drift rate of the hydrogen maser as inferred by the ratio between the OLC and the hydrogen maser during each measurement period. 

To estimate the uncertainty associated with stochastic fluctuations of the maser, we adopt a Monte-Carlo approach based on simulating month-long hydrogen maser time series using a model of its frequency noise. For each time series, the offset between the mean frequency during the uptime period $T_{\mathrm{UP}}$ and the entire time series $T_{\mathrm{ALL}}$ is calculated. The standard deviation of these simulated offsets provides an uncertainty estimate for possible frequency errors arising from the stochastic noise and downtime.

The noise model for the hydrogen maser is based on comparisons against the OLC and is shown in figure \ref{fig:maser}b. It is important to have several extended periods of measurement to properly capture the behaviour of the maser over long periods of measurement downtime. For this campaign, the OLC was operational for 74\% of the time with several continuous stretches exceeding 24 hours. To reveal the flicker floor of the maser at long time scales we relaxed our uptime criteria to overlook downtime periods of up to 2500 seconds, which yielded several quasi-continuous periods extending over several days. We found that a noise model comprised of white phase noise averaging down as $4 \times 10^{-13}\,\tau^{-1}$, combined with white frequency noise averaging down as $12 \times 10^{-14} \,\tau^{-1/2}$ and flicker noise at the $8 \times 10^{-16}$ level, closely approximated the measured maser instability for all observed time scales. Using this model, we simulated one hundred time series extending over both measurement periods using Allantools---an open source software package developed for python \cite{allantools}. Following this procedure, we estimate the uncertainty contributed by the combination of stochastic maser noise and measurement downtime to be \num{1.4e-16} and \num{2.5e-16} for the two measurement periods. 

%\begin{tabular}{lccccccc}
%\hline\hline
%Partial Ratio & Contribution & $r_0$ &  Period 1  & Period 1 &  Period 2  & Period 2 \\
% &  & $r_0$ & $ r/r_0 - 1$  & u$[r/r_0 - 1]$ \\
%\hline
%f(Sr)/f(UTC(NPL); T$_{\mathrm{UP}}$) & Ratio at Comb & 429 228 004 229 873  & -5542 &--&\\
%  &Sr Statistical&& -- & 10 \\
%\hline
%f(UTC(NPL); T$_{\mathrm{UP}}$)/f(UTC(NPL)) & Deterministic & 1 & -223 & 5\\
%& Stochastic&  & -- &200\\
%\hline
%f(UTC(NPL))/f(TAI) & &  1& 2320 & 640 \\
%\hline 
%f(TAI))/f(SI) & &  1 & -70 & 210 \\
%\hline
%Systematic Corrections & Sr Systematics &  4292 280 042 29 873 & -70 & 210\\
%                       & Gravity &   & -70 & 210\\
%\hline
%$\frac{f_{\mathrm{Sr}}}{f_{\mathrm{SI}}}$ & &  429 228 004 229 873 & 438  & 703 \\             %\hline\hline                  
%\end{tabular}
%\end{table*}

\begin{table*}
\centering
\caption{\label{tab:absfreq} Summary of each of the ratios $r$ specified in equation \ref{eq:ratioEquationMod} which are combined to compute the absolute frequency of the transition. For each ratio, the fractional deviation from its nominal value $r_0$ and corresponding uncertainty (68\% confidence interval) is reported. Fractional values are in units of $1 \times 10^{-18}$. The value \textit{R}~=~\num{429228004229873.0} is based on the 2017 CIPM recommended frequency value for the 5s$^2$ $^1$S$_0$ to 5s5p $^3$P$_0$ clock transition in $^{87}$Sr.}

\begin{tabular}{lccccccc}
\toprule
Ratio & Contribution & $r_0$ &  Period 1  & Period 1 &  Period 2  & Period 2 \\
 &  &  & $ r/r_0 - 1$  & u$[r/r_0 - 1]$ & $ r/r_0 - 1$  & u$[r/r_0 - 1]$ \\
\hline
$f_{\mathrm{Sr}}/f_{\mathrm{UTC(NPL);T_\mathrm{UP}}}$ & Ratio at comb & \textit{R}  & -3917 &100&  -4437  &100&\\
  &Sr statistical&& -- & 2 & -- & 3.5 \\
 & Sr systematics &   & 5146 & 10 & 5185& 10  \\
                       & Gravity &   & -1214 & 4 &-1214 & 4\\
\hline
$f_{\mathrm{UTC(NPL);T_\mathrm{UP}}}/f_{\mathrm{UTC(NPL);T_{ALL}}}$  & Deterministic & 1 & 216  & 50 & 119  & 30\\
& Stochastic&  & -- &140& -- &250\\
\hline
$f_{\mathrm{UTC(NPL);T_{ALL}}}/f_{\mathrm{TAI;T_{ALL}}}$ & Local time scale to TAI &  1&  154 & 1200 &  463 & 3200 \\
\hline 
$f_{\mathrm{TAI;T_{ALL}}}/f_{\mathrm{SI}}$ & TAI to SI second &1&  -200 & 370 & 90 & 710 \\

\hline

$f_{\mathrm{Sr}}/f_{\mathrm{SI}}$ & Total &  \textit{R}  & 185  & 1300 & 206  & 3300\\              
\bottomrule
\end{tabular}
\end{table*}

\begin{figure}[t]
    \centering
    \includegraphics[width=0.48\textwidth]{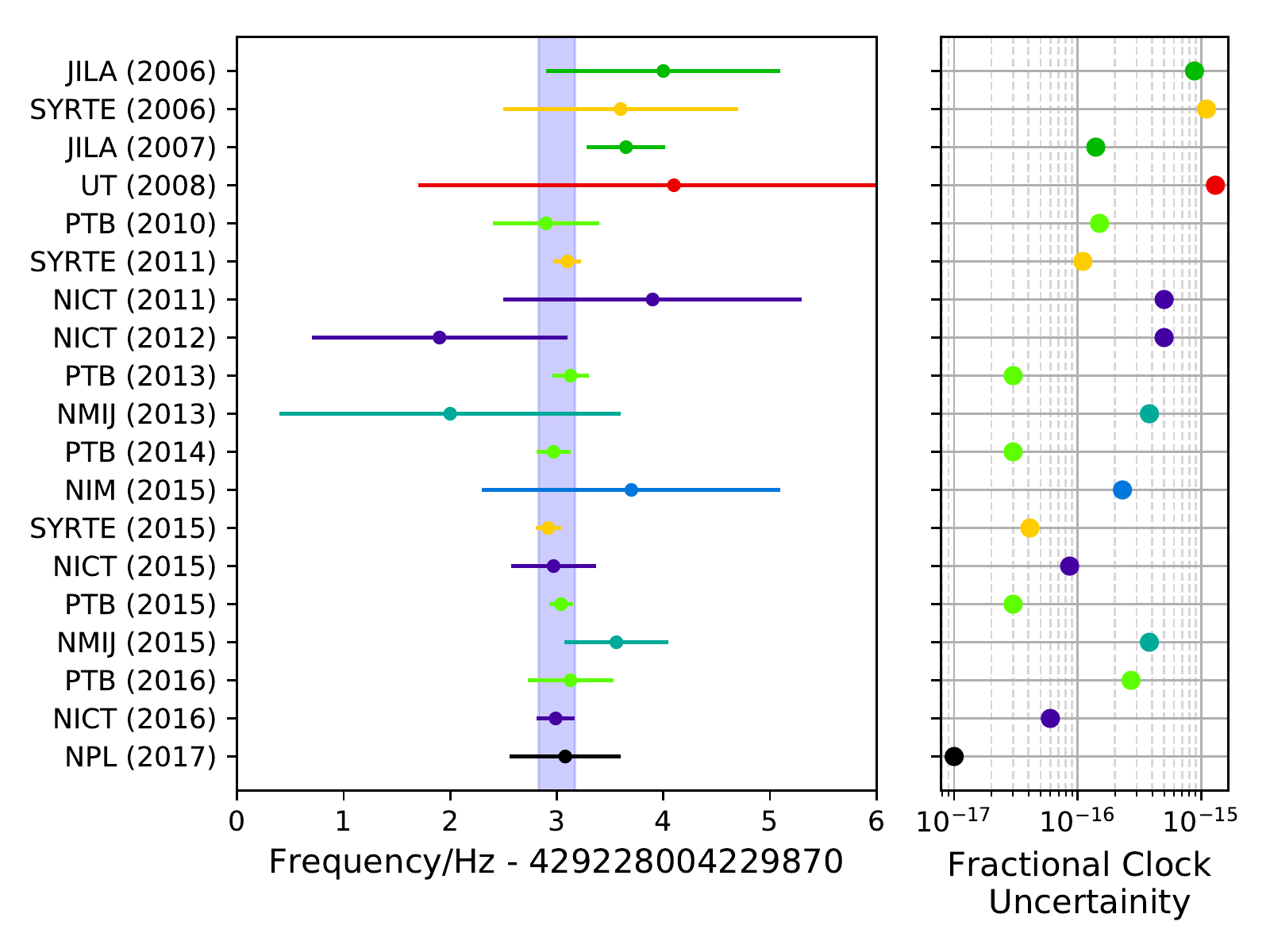}
    \caption{Left: Summary of all absolute frequency measurements of the 5s$^2$~$^1$S$_0$ to 5s5p~$^3$P$_0$ clock transition in $^{87}$Sr since the CIPM first recommended its value in 2006. Measurements were recorded at JILA (\protect\tikz\fill[color={rgb:red,.36;green,1;blue,0.0}] (0,0) circle (.6ex);) \cite{Boyd2007b, Campbell2008}, University of Tokyo (\protect\tikz\fill[color={rgb:red,.5;green,0;blue,0}] (0,0) circle (.6ex);) \cite{Hong2009}, SYRTE (\protect\tikz\fill[color=yellow] (0,0) circle (.6ex);) \cite{Baillard2007, LeTargat2013, Lodewyck2016}, PTB (\protect\tikz\fill[color={rgb:red,0;green,.73;blue,0}] (0,0) circle (.6ex);) \cite{Falke2011b, Falke2014, Grotti2018, Koller2017}, NICT \cite{Matsubara2009, Yamaguchi2011, Hachisu2016, Hachisu2017} (\protect\tikz\fill[color={rgb:red,.27;green,0;blue,0.63}] (0,0) circle (.6ex);), NMIJ \cite{Akamatsu2014, Tanabe2015} (\protect\tikz\fill[color={rgb:red,0;green,.67;blue,0.6}] (0,0) circle (.6ex);), NIM \cite{Lin2015} (\protect\tikz\fill[color={rgb:red,0;green,.46;blue,0.87}] (0,0) circle (.6ex);), and NPL (\protect\tikz\fill[color=black] (0,0) circle (.6ex);). Also shown is the updated value for the transition frequency as recommended by the CIPM in 2017 (blue-shaded region) \cite{Riehle2018}. Right: Contribution from the systematic uncertainty of the strontium clocks -- neglecting gravitational redshift -- to the total uncertainty of each absolute frequency measurement. %Not included are results reported at JILA \cite{Nicholson2015} and Riken \cite{Ushijima2015}---which both have predicted uncertainties below those presented above---as no absolute frequency measurement exists for these systems. %However for the clocks at Riken, two independent systems were compared and showed agreement at the mid-$10^{-18}$ level.
    \label{fig:history}}
\end{figure}

\subsection{Local time scale to TAI: $f_{\mathrm{UTC(NPL);T_{all}}}/f_{\mathrm{TAI;T_{all}}}$}

The time offset between UTC(NPL) and UTC is computed by the BIPM at
5-day intervals and published in the monthly Circular T bulletin. As our measurement period aligns with these 5-day intervals, the accumulated time offset can be used to compute the mean frequency difference between our local time scale and TAI over the measurement period. The fractional uncertainty associated with this offset is calculated based on the type-A link timing uncertainty as specified in Circular T, which for this campaign month was 1~ns. To account for correlations between measurements, we do not directly divide the uncertainty by the total measurement duration \textit{T}, but instead extrapolate the error as \cite{Panfilo2010}: 

\begin{equation}
u\left[\frac{f(\mathrm{UTC(NPL))}}{f(\mathrm{TAI})}-1\right] = \frac{\sqrt{2}\times1\,\mathrm{ ns}}{86400\, \mathrm{s}\times5}\left(\frac{5}{T}\right)^{0.9}.
\end{equation}

\subsection{TAI to the SI second: $f_{\mathrm{TAI;T_{all}}}/f_{\mathrm{SI}}$}

To complete the evaluation of the secondary frequency standard against the SI second, the average deviation \textit{d} over the measurement period of the scale interval of TAI from the SI second was calculated. %\st{To complete the evaluation of the secondary frequency standard against the SI second the fractional frequency deviation of TAI against the SI second on the rotating geoid during the measurement must be calculated.}
 The computation of the \textit{d} values for the two measurement periods was carried out by the BIPM using the same algorithm and data that produces the \textit{d} values reported in Circular T for each one-month interval of TAI. For the first and second measurement intervals the \textit{d} values were $2.0(37) \times10^{-16}$ and $-0.9(71) \times10^{-16}$, respectively. 
 
We also account for general-relativistic effects by transforming from the proper time of the clock to TAI. The relativistic rate shift is computed with respect to the conventionally adopted equipotential $W_0$~=~\SI{62636856.0}{\meter\squared\per\second\squared} of the Earth’s gravity potential. As previously reported in \cite{riedel2020}, this correction was determined to be $-12.14(4)\times10^{-16}$ following the procedure outlined in \cite{Denker2018} as part of the EMRP project international timescales with optical clocks (ITOC) \cite{Margolis2013}. The final absolute frequency measurements for both intervals are summarized in table \ref{tab:absfreq} which also outlines the correction and associated uncertainty contributed by each frequency ratio in equation \ref{eq:ratioEquationMod}. Combining the results of the two measurements, we estimate the absolute frequency of the OLC to be \num{429228004229873.1}(5)~Hz. This is in good agreement with the 2017 CIPM recommended frequency value of \num{429228004229873.0}~Hz, which has a fractional uncertainty of \num{4e-16} \cite{Riehle2018}, as well as with measurements at other institutes, as shown in figure \ref{fig:history}. Figure \ref{fig:history} also shows the systematic uncertainties of the OLCs at the time that their absolute frequencies were reported. Note that the JILA \cite{bothwell2019} and Tokyo \cite{Ushijima2015} groups have since improved their clocks and both now have predicted uncertainties at the low $10^{-18}$ range.
 
\section{Conclusion}

We have realised a strontium optical lattice clock with an estimated systematic uncertainty of \num{1e-17}, and determined its absolute frequency by comparison against TAI to be \num{429228004229873.1}(5)~Hz. Future work will focus on direct comparison against other optical clocks with a precision at or below the \num{1e-17} level. As part of an optical fibre network linking us to other optical clocks around Europe, we will take part in long-distance comparisons to verify clock accuracy and test relativistic physics \cite{Delva2017,Lisdat2015}. To prepare for such measurements, several improvements to the clock performance are underway: an improved \SI{48.5}{\centi\meter} ULE cavity, similar to the design in \cite{Hafner2015}, is expected to significantly improve our frequency instability, while the uncertainty in BBR and lattice shifts will be reduced by introducing an air gap between the MOT coils and the chamber, and by applying a more complete model of the lattice potential \cite{Brown2017, Ushijima2018}.

\section{Acknowledgements}

The authors thank G\'erard Petit for calculating \textit{d} values for our custom measurement intervals, Rachel Godun and Peter Whibberley for helpful discussions, and Karen Alston and Radka Veltcheva for temperature sensor calibration. We also note that our absolute frequency measurement derives its accuracy from primary and secondary standards operated at other national measurement institutes around the world.

This work was financially supported by: the UK Department for Business, Energy and Industrial Strategy as part of the National Measurement System Programme; the European Metrology Research Programme (EMRP) project SIB55-ITOC; and the European Metrology Programme for Innovation and Research (EMPIR) project 15SIB03-OC18. This work has received funding from the EMPIR programme co-financed by the Participating State and from the European Union\rq s Horizon 2020 research and innovation programme. The EMRP is jointly funded by the EMRP participating countries within EURAMET and the European Union.  W.B. would like to acknowledge the EU Innovative Training Network (ITN) Future Atomic Clock Technology (FACT).

\section*{References}
\bibliographystyle{iopart-num}
\bibliography{refs}

\end{document}